\documentclass[aps,prd,onecolumn,groupedaddress,showpacs,nofootinbib,amssymb]{revtex4-2}
\usepackage[dvips]{graphicx}
\usepackage{amssymb}
\usepackage{amsmath}
\usepackage{graphicx,,color}
\usepackage{amsfonts}
\usepackage{bm}
\usepackage{cancel}
\usepackage{comment}
\usepackage{floatflt}
\newcommand\be{\begin{equation}}
\newcommand\ee{\end{equation}}

\allowdisplaybreaks[4]

\begin{document}

\title{Solving the Trans-Planckian Censorship Problem with a Power-law Tail in $R^2$ Inflation: A Dynamical System Approach}
\author{S.D. Odintsov$^{1,2,3}$}\email{odintsov@ice.csic.es}
\author{V.K. Oikonomou,$^{3,4}$}\email{voikonomou@gapps.auth.gr;v.k.oikonomou1979@gmail.com}
\author{Eleni I. Manouri,$^{4}$}\email{elenimanouri21@gmail.com;emanouri@physics.auth.gr}
\author{Asterios T. Papadopoulos,$^{4}$}\email{apapadopb@auth.gr;asterispap05@gmail.com}
\affiliation{$^{1)}$ Institute of Space Sciences (ICE, CSIC) C. Can Magrans s/n, 08193 Barcelona, Spain \\
$^{2)}$ ICREA, Passeig Luis Companys, 23, 08010 Barcelona, Spain\\
$^{3)}$L.N. Gumilyov Eurasian National University - Astana,
010008, Kazakhstan\\
$^{4)}$Department of Physics, Aristotle University of
Thessaloniki, Thessaloniki 54124, Greece}

%$^{2)}$ Laboratory for Theoretical Cosmology, International Center
%of Gravity and Cosmos, Tomsk State University of Control Systems
%and Radioelectronics  (TUSUR), 634050 Tomsk, Russia

 \tolerance=5000

\begin{abstract}
In this work we elaborate on solving the trans-Planckian
censorship problem of standard slow-roll inflation by using a
power-law inflationary tail generated by a scalar field with an
exponential potential. We use a quantitative approach by studying
in detail the phase space of a combined $F(R,\phi)$ cosmological
system, focusing on the de Sitter and power-law subspaces of the
total phase space. As we show, the de Sitter subspace of the
$F(R,\phi)$ system shares the same fixed points as the vacuum
$F(R)$ gravity system and the trajectories in the phase space tend
to these fixed points. However, the power-law subspace is not
stable and cannot be realized by the combined $F(R,\phi)$ system.
To this end, we propose a well-motivated phenomenological $F(R)$
gravity model for which the $R^2$ term is switched off below a
critical curvature near the end of the $R^2$ slow-roll
inflationary era, and below that critical curvature, only the
Einstein-Hilbert gravity term and the scalar field remain in the
effective inflationary Lagrangian. The remaining system can
successfully realize a power-law tail of the $R^2$ slow-roll era.
\end{abstract}

%PACS numbers: 04.50.Kd, 95.36.+x, 98.80.-k, 98.80.Cq

\maketitle

\section{Introduction}

Understanding the primordial era of our Universe is one of the
most prominent problems in modern theoretical physics. The most
appealing and self-consistent description of the primordial era is
inflation \cite{inflation1,inflation2,inflation3,inflation4} which
remedies theoretically the shortcomings of the Standard Big Bang
theory. The inflationary era belongs to the classical era of our
Universe, means that the spacetime is four dimensional during
inflation and the quantum modes that exit the Hubble horizon
during the inflationary era, freeze and become classical after
they exit the horizon. This is a standard requirement for
inflationary theories, which shall concern us in this paper.
Inflation can be realized in various contexts, for example in the
context of General Relativity (GR) one can use a scalar field,
minimally or non-minimally coupled with gravity
\cite{inflation1,inflation2,inflation3,inflation4} but also
modified gravity can realize the inflationary era, with the most
prominent modified gravity theory
\cite{reviews1,reviews2,reviews3,reviews4} being $F(R)$ gravity
\cite{Nojiri:2003ft,Capozziello:2005ku,Hwang:2001pu,Cognola:2005de,Song:2006ej,Faulkner:2006ub,Olmo:2006eh,Sawicki:2007tf,Faraoni:2007yn,Carloni:2007yv,
Nojiri:2007as,Deruelle:2007pt,Appleby:2008tv,Dunsby:2010wg}, for
various phenomenological and mathematical reasons. There are
various motivations for using modified gravity to describe the
Universe's evolution, with the most strong motivation coming from
the late-time cosmology. In the context of GR, the late-time
evolution can be described by a cosmological constant, however the
cosmological constant description of the dark energy era is not
compatible with the latest DESI 2025 data \cite{DESI:2025zgx}
which point out that the dark energy is dynamical at an impressive
4.2$\sigma$ statistical confidence. Also even the Planck data
indicate that the dark energy era might be described by a phantom
equation of state (EoS) which in the context of GR can be realized
by tachyon fields. Although tachyons are theoretically motivated
by string theory, a consistent description of our Universe should
avoid such scalar fields. Thus modified gravity seems to be a
consistent proposal for the description of our Universe, and in
fact it is possible that a unified description of inflation and
the dark energy era can be provided by modified gravity and
specifically $F(R)$ gravity, see for example the pioneer article
\cite{Nojiri:2003ft} on this topic.

Inflation will be in the focus of many future experiments and
collaborations, like the Simons observatory
\cite{SimonsObservatory:2019qwx} and the future gravitational wave
experiments
\cite{Hild:2010id,Baker:2019nia,Smith:2019wny,Crowder:2005nr,Smith:2016jqs,Seto:2001qf,Kawamura:2020pcg,Bull:2018lat,LISACosmologyWorkingGroup:2022jok}.
The Simons observatory will probe directly the $B$-modes in the
Cosmic Microwave Background (CMB) while the gravitational wave
experiments will seek for a stochastic gravitational wave
background generated by the inflationary era. In 2023 NANOGrav and
other Pulsar Timing Arrays collaborations
\cite{nanograv,Antoniadis:2023ott,Reardon:2023gzh,Xu:2023wog}
already reported the existence of a stochastic gravitational wave
background, but it is highly unlikely that this background is due
to inflation \cite{Vagnozzi:2023lwo,Oikonomou:2023qfz}.

Now, the question is which theory describes optimally the
inflationary era. The answer is not easy, because there is
motivation for having scalar fields and higher curvature terms in
the inflationary effective Lagrangian. Hence, it might be possible
that one cannot use one of the two descriptions, but one must
include both scalars and higher order curvature invariants in the
effective inflationary Lagrangian. The reason is simple and it is
highly motivated by the quantum action of a scalar field.
Specifically, the scalar field in its vacuum configuration can
have at tree order the following action,
\begin{equation}\label{generalscalarfieldaction}
\mathcal{S}_{\varphi}=\int
\mathrm{d}^4x\sqrt{-g}\left(\frac{1}{2}Z(\varphi)g^{\mu
\nu}\partial_{\mu}\varphi
\partial_{\nu}\varphi+\mathcal{V}(\varphi)+h(\varphi)R
\right)\, ,
\end{equation}
and can be either conformally coupled or minimally coupled. The
quantum corrected effective action of the scalar field in its
vacuum configuration contains the following terms
\cite{Codello:2015mba},
\begin{align}\label{quantumaction}
&\mathcal{S}_{eff}=\int
\mathrm{d}^4x\sqrt{-g}\Big{(}\Lambda_1+\Lambda_2
R+\Lambda_3R^2+\Lambda_4 R_{\mu \nu}R^{\mu \nu}+\Lambda_5 R_{\mu
\nu \alpha \beta}R^{\mu \nu \alpha \beta}+\Lambda_6 \square R\\
\notag & +\Lambda_7R\square R+\Lambda_8 R_{\mu \nu}\square R^{\mu
\nu}+\Lambda_9R^3+\mathcal{O}(\partial^8)+...\Big{)}\, ,
\end{align}
and we included terms which make the action compatible with
diffeomorphism invariance and also that terms contain up to
fourth-order derivatives. Note that the parameters $\Lambda_i$,
$i=1,2,...,6$ are dimensionful constants. Hence, a complete
description of the effective inflationary Lagrangian might contain
a scalar field along with higher order curvature invariants.

In a recent work, we used the combination of having a scalar field
with exponential potential in the presence of an $R^2$ term in the
inflationary Lagrangian as a possible solution of the
trans-Planckian problem for the inflationary modes. We showed that
if the slow-roll inflationary era, generated by an $R^2$ term, is
followed by a power-law inflationary tail, generated by a scalar
field with an exponential potential, the trans-Planckian issues of
standard slow-roll inflation find a self-consistent remedy
\cite{transplanckOdintsovOikonomou}. In the literature, having
$R^2$ terms in the presence of a scalar field is quite common
\cite{Ema:2017rqn,Ema:2020evi,Ivanov:2021ily,Gottlober:1993hp,Enckell:2018uic,Kubo:2020fdd,Gorbunov:2018llf,Calmet:2016fsr,Oikonomou:2021msx,Oikonomou:2022bqb},
but in our approach, the scalar field has an exponential
potential, which generates a power-law tail at the end of the
inflationary era. However our previous study
\cite{transplanckOdintsovOikonomou} was a qualitative approach
based on the arguments that a power-law tail would actually remedy
the Trans-Planckian Censorship Conjecture (TCC)
\cite{Martin:2000xs,Brandenberger:2000wr,Bedroya:2019snp,Brandenberger:2021pzy,Brandenberger:2022pqo,Kamali:2020drm,Berera:1995ie,Brandenberger:2025hof}
issues of standard slow-roll inflation, if the scalar field
dominates the evolution near the end of a standard $R^2$ slow-roll
era. The question however is whether such a scenario can be
realized by the cosmological system of an $R^2$-corrected
exponential scalar theory. This question can be concretely
answered if the phase space of the system is studied in detail. In
this article, we perform a dynamical system analysis of the
combined $R^2$ corrected scalar theory with the scalar having an
exponential potential. We construct an autonomous dynamical system
using appropriate dimensionless variables and we analyze in depth
the phase space of the combined system, focusing on de Sitter and
power-law subspaces of the total phase space. As we demonstrate,
the de Sitter subspace is controlled by the $R^2$ gravity, since
the trajectories in the phase space tend to the same fixed points
that the vacuum $F(R)$ gravity has. This proves that indeed the
$R^2$ gravity will realize a quasi-de Sitter era before the
power-law inflationary era. However, the power-law subspace is
quite unstable and it is highly unlikely that it can be physically
realized by the combined $R^2$-scalar field system. Thus in order
to solve the TCC issues of the standard slow-roll inflationary era
generated by the $R^2$ gravity, one must use a phenomenological
$F(R)$ gravity for which, below a critical curvature, the $R^2$
gravity is switched off and only Einstein-Hilbert gravity with a
scalar field remains. This system can realize a pure power-law
evolution, producing the desirable power-law tail that can solve
the TCC problems of $R^2$ inflation. The phenomenological model of
$F(R)$ gravity we will use is an exponential deformation of
standard $R^2$ gravity, motivated by the positivity of the de
Sitter perturbation scalaron mass \cite{modelagnosticFR}.

This article is outlined as follows: In section II we review the
mechanism of how a power-law tail in the standard quasi-de Sitter
slow-roll $R^2$ era can solve the TCC problems of the vacuum $R^2$
gravity. In section III we study in depth the complete phase space
of the $R^2$-corrected scalar field theory with exponential
potential. By using appropriate dimensionless parameters we
construct an autonomous dynamical system and we study its quasi-de
Sitter and power-law phase spaces. As we show the de Sitter
subspace is controlled by the $R^2$ gravity, however the power-law
subspace is highly unstable and we show that it is highly unlikely
that this can be physically realized. In section IV we propose a
phenomenological $F(R)$ gravity model, which below a critical
curvature near the end of the inflationary era, can switch off the
$R^2$ term. We show that the remaining scalar system can generate
the desired power-law tail of the $R^2$ era.

Before proceeding, in this work we shall use a flat
Friedmann-Robertson-Walker (FRW) geometric background, with the
line element being,
\begin{equation}\label{frw}
ds^2 = - dt^2 + a(t)^2 \sum_{i=1,2,3} \left(dx^i\right)^2\, ,
\end{equation}
where $a(t)$  is as usual the scale factor and also the Ricci
scalar for the FRW metric is,
\begin{equation}\label{ricciscalaranalytic}
R=6\left (\dot{H}+2H^2 \right )\, ,
\end{equation}
where $H=\frac{\dot{a}}{a}$ stands for the Hubble rate.

\section{Overview of $f(R,\phi)$ Gravity Inflationary Framework and TCC Modes with Power-law Tail of $R^2$ Inflation}

In this section we shall briefly provide a concrete overview of
our previous work \cite{transplanckOdintsovOikonomou} and discuss
the essential features of the idea that a power-law tail of a
standard slow-roll $R^2$ inflationary theory can provide a remedy
for the TCC problems of standard slow-roll inflation. We will base
the presentation on our previous work Ref.
\cite{transplanckOdintsovOikonomou} in order to provide the
correct context for the next sections of this article. As we
mentioned in the introduction, the standard slow-roll inflationary
era, realized by minimally coupled scalar field theory or $R^2$
gravity, or any other modified gravity theory, has a serious issue
related to the TCC
\cite{Martin:2000xs,Brandenberger:2000wr,Bedroya:2019snp,Brandenberger:2021pzy,Brandenberger:2022pqo,Kamali:2020drm,Berera:1995ie,Brandenberger:2025hof}.
The TCC indicates that quantum inflationary modes with wavelengths
that are trans-Planckian must not exit the Hubble horizon during
the inflationary era and thus remain forever quantum and never
enter the classical regime of modes outside the Hubble horizon.
This TCC requirement for standard slow-roll inflationary regimes
would impose severe conditions on the scale of inflation at the
end of inflation which should be $V_e<10^{10}\,$GeV, which in turn
imposes constraints on the tensor-to-scalar ratio to be
$r<10^{-30}$. Apparently these two constraints eliminate most of
the Planck-2018-compatible inflationary theories. In our previous
work Ref. \cite{transplanckOdintsovOikonomou} however, we found a
remedy for the TCC problems of standard slow-roll inflation.
Recall that during the inflationary era, the Hubble horizon
shrinks in a nearly inverse exponential way, due to the quasi-de
Sitter evolution, and shrinks up to the point that the slow-roll
condition breaks down, so when the first slow-roll index becomes
of the order of unity. Now our approach in Ref.
\cite{transplanckOdintsovOikonomou} was simple, we assumed that
before the end of the $R^2$ slow-roll inflationary era, a scalar
field with exponential potential dominates the evolution. Such a
framework is known to produce a constant EoS evolution, as we now
briefly show for completeness. Specifically, the scalar field is
assumed to satisfy the following constraint,
\begin{equation}\label{eoscondition}
\dot{\phi}^2=\beta V(\phi)\, ,
\end{equation}
hence,
\begin{equation}\label{dddotphi}
\ddot{\phi}=\frac{\beta V'}{2}\, .
\end{equation}
The scalar field equation of motion is,
\begin{equation}\label{fieldfree}
\ddot{\phi}+3H\dot{\phi}+V'=0\, ,
\end{equation}
hence we have,
\begin{equation}\label{eqnfreescalarkin}
\left(\frac{\beta+2}{2} \right)^2\left(
V'\right)^2=9H^2\dot{\phi}^2\, ,
\end{equation}
or equivalently,
\begin{equation}\label{potentialapprox}
V=V_0e^{-\sqrt{\frac{6\beta}{\beta+2}}\kappa \phi}\, .
\end{equation}
The exponential potential of Eq. (\ref{potentialapprox}) is
exactly the form of the potential we shall use in this work. In
the following we define the parameter $\lambda$ in the following
way,
\begin{equation}\label{lambda}
\lambda=\sqrt{\frac{6\beta}{\beta+2}}\, ,
\end{equation}
therefore the scalar field potential is written as,
\begin{equation}\label{exponentialpotential}
V(\phi)=V_0\,e^{-\lambda \phi \kappa}\, .
\end{equation}
Now such an evolution for the scalar field yields the following
total EoS, if the evolution is scalar field dominated,
\begin{equation}\label{eosscalarfinal}
w_{\phi}=\frac{\beta-2}{\beta+2}\, .
\end{equation}
The above evolution describes an inflationary power-law tail with
a scale factor $a(t)=a_0\,t^{\eta}$, with $\eta=\frac{\beta +2}{3
\beta }$, if $\beta<1$ and in the following we shall use the
choice $\beta=0.99$ as we did in Ref.
\cite{transplanckOdintsovOikonomou}. The way that the power-law
inflationary tail of the $R^2$ era works is conceptually as
follows: Following the arguments of Ref.
\cite{transplanckOdintsovOikonomou,Kamali:2020drm}, the energy
density of the scalar field after the end of the slow-roll
inflationary era and when the power-law inflationary tail starts,
is,
\begin{equation}\label{b1}
\rho_{\phi}\sim a^{-\frac{3\beta}{1+\frac{\beta}{2}}}\, .
\end{equation}
The spatial flatness requirement is to
\cite{transplanckOdintsovOikonomou,Kamali:2020drm},
\begin{equation}\label{b3}
\left(\frac{a_i}{a_R}\right)^{2-\frac{3\beta}{1+\frac{\beta}{2}}}<10^{-2}\frac{T_0T_{eq}}{T_R^2}\,
,
\end{equation}
while in terms of the scalar field evolution and potential we have
\cite{transplanckOdintsovOikonomou,Kamali:2020drm},
\begin{equation}\label{b10}
\left( \frac{a_i}{a_R}
\right)^{2-\frac{3\beta}{\sqrt{1+\frac{\beta}{2}}}}>\frac{T_0T_{eq}}{T_R^2}\,
,
\end{equation}
where $a_i$ is the value of the scale factor at the beginning of
the inflationary power-law tail, $a_R$ is the scale factor at the
end of the power-law inflationary tail, $T_0$ is the temperature
of the Universe today, $T_{eq}$ is the temperature of the Universe
at matter-radiation equality, and $T_R$ is the temperature of the
Universe at the end of the power-law tail. Now when $\beta=0$ or
equivalently $w_{\phi}\sim -1$, the inequality of Eq. (\ref{b10})
would be in conflict with that of Eq. (\ref{b3}). However, for
larger values of $\beta$ in the range $0.5<\beta<1$ the two
inequalities are intact. We will take $\beta\sim 0.99$ which
yields a scalar field EoS parameter $w_{\phi}\sim -0.337793$. The
TCC is solved as follows, the censorship itself is embodied in the
following inequality,
\begin{equation}\label{tcc}
\frac{a_R}{a_i}\ell_{pl}<H^{-1}(t_R)\, ,
\end{equation}
where $\ell_{pl}$ is the Planck length. So following
\cite{transplanckOdintsovOikonomou,Kamali:2020drm} we get the
resulting expression for the condition that is required to hold
true if the TCC must be respected
\cite{transplanckOdintsovOikonomou,Kamali:2020drm},
\begin{equation}\label{tr}
\frac{T_R}{M_p}<\left( 3/
g^*\right)^{(1-\tilde{\beta})/(6-4\tilde{\beta})}\times10^{-2/(6-4\tilde{\beta})}\left(\frac{T_0T_{eq}}{M_p}
\right)^{1/(6-4\tilde{\beta})}\, ,
\end{equation}
where $\tilde{\beta}=\frac{3\beta}{\beta+2}$. Since in the case we
are considering, we have $\beta=0.99$, we get, $T_R<10^{-30}\times
M_p$, but this does not affect the flatness issue, which was
solved due to the inequality (\ref{b10}) and due to the fact that
an $R^2$ era precedes the power-law tail for a large number of
$e$-foldings. Also such a low temperature at the end of the
power-law tail does not affect the scalar perturbations or the
tensor perturbations of the CMB because the power-law tail is
detached from the $R^2$ slow-roll era. This is due to the fact
that during the power-law tail the Hubble horizon shrinks in a
much slower rate compared to the nearly inverse de Sitter Hubble
horizon shrinking. Thus, the power-law tail also contributes to
the CMB, but possibly these modes have wavelengths much shorted
than 10Mpc, hence although these modes exit from the Hubble
horizon during the power-law inflationary tail, do not affect the
CMB at least linearly. At these wavelengths the CMB is non-linear.
\begin{figure}[h]
\centering
\includegraphics[width=45pc]{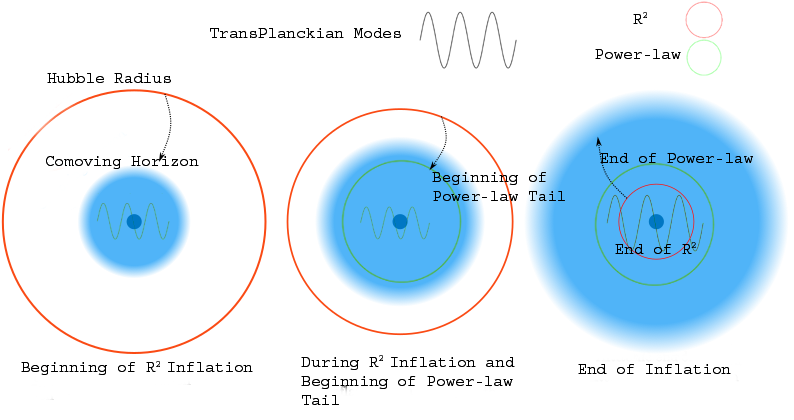}
\caption{Schematic presentation of the resolution of TCC problems
offered by the power-law inflationary tail following a standard
$R^2$ slow-roll inflationary era. When the $R^2$ slow-roll
inflationary era commences (red circle, left figure) the Hubble
horizon is large and all the inflationary modes are contained in
it. As the slow-roll inflationary era proceeds, the inflationary
modes gradually exit the Hubble horizon and become classical
(right figure, red circle). If no power-law era follows right
after the $R^2$ slow-roll era, the Trans-Planckian modes would
also exit the horizon (red circle, right figure), hence we invoked
an inflationary power-law tail that follows the standard $R^2$
slow-roll era (green circle, figure at the center). During the
power-law inflationary tail, the Hubble horizon shrinks more
slowly compared to the de Sitter shrinking of the $R^2$ era,
therefore the Trans-Planckian modes never exit the Hubble horizon
and never become classical (green circle, right figure). Hence,
this slowing down of the shrinking rate of the Hubble horizon
during the power-law inflationary tail is what solves the TCC
problems of standard slow-roll inflation.} \label{plot1}
\end{figure}
Let us illustrate schematically our way of thinking on how a
power-law inflationary tail might solve the TCC problems of
standard $R^2$ slow-roll inflation. In Fig. \ref{plot1} we present
the various steps of the resolution we proposed in Ref.
\cite{transplanckOdintsovOikonomou} in a schematic way. Let us
describe the procedure that is depicted in Fig. \ref{plot1}. At
the beginning of the $R^2$ slow-roll inflationary era the Hubble
horizon (red circle, left figure) is large and all the
inflationary modes are contained in it. As the slow-roll
inflationary era proceeds, more and more modes exit the Hubble
horizon and become classical (right figure, red circle). If no
power-law era follows the $R^2$ slow-roll era, the trans-Planckian
modes would also exit the horizon (red circle, right figure), so
we invoked an inflationary power-law tail to follow the $R^2$ era
(green circle, figure at the center). Now during the power-law
inflationary tail, the Hubble horizon shrinks more slowly compared
to the de Sitter shrinking of the $R^2$ era, thus the
trans-Planckian modes never exit the Hubble horizon (green circle,
right figure) and thus never become classical. Hence, this slowing
down of the shrinking rate of the Hubble horizon during the
power-law inflationary tail, is what solves the TCC problems of
slow-roll inflation. Of course, it is notable that several
inflationary modes of the power-law tail inflationary era will
contribute to the CMB. However, since the shrinking rate of the
power-law tail is small, it is possible that the modes that exit
the horizon have a wavelength smaller than $10$Mpc, thus these
modes contribute to the non-linear parts of the CMB.

Now the important question we asked in our previous work
\cite{transplanckOdintsovOikonomou} but we did not address
formally, is whether our scenario is realized by the combined
field equations of the $R^2$-scalar field system, since we
required a clean power-law inflationary tail in which the $R^2$
gravity will not dominate the evolution. This study is the core
part of this work and is presented in the next sections.

\section{Study of the Complete $F(R,\phi)$ Gravity Phase Space for de Sitter and Power-law Cases}

In this section we will study the complete phase space of the
$F(R,\phi)$ gravity system. We will focus on the case that the
parameter $\beta$ in the scalar field subsystem is $\beta=0.99$
and we examine the phase space using an autonomous dynamical
system that we will extract using the field equations. Our aim is
to see whether the total phase space of the system contains the
power-law tail controlled by the scalar field at the end of the
$R^2$ inflation. Also we shall answer the question whether the
$R^2$ term drives the quasi-de Sitter evolution of the combined
scalar-$F(R)$ gravity system. Our results are striking since it
proves that the power-law cosmology is never realized by the
$F(R,\phi)$ gravity system, but we also prove that the $R^2$ term
indeed realizes a quasi-de Sitter era for the combined
scalar-$F(R)$ gravity system. In fact, as we will show the only
inflationary evolution that can be realized by the $F(R,\phi)$
gravity system is the de Sitter evolution. Let us get into the
details of our analysis to present concretely these striking
results.

We start with the combined $F(R,\varphi)$ gravitational action
which has the form,
\begin{equation}
\label{action} \centering
\mathcal{S}=\int{d^4x\sqrt{-g}\left(\frac{
F(R)}{2\kappa^2}-\frac{1}{2}g^{\mu \nu}\partial_{\mu}\varphi
\partial_{\nu}\varphi-\mathcal{V}(\varphi)\right)}\, ,
\end{equation}
with $\kappa^2=8\pi G=\frac{1}{Mp^2}$ and also $M_p$ stands for
the reduced Planck mass. Varying the gravitational action with
respect to the metric tensor and with respect to the scalar field,
we obtain the following field equations,
\begin{align}\label{eqnsofmkotion}
& 3 H^2F_R=\frac{RF_R-F}{2}-3H\dot{F}_R+\kappa^2\left(
\frac{1}{2}\dot{\phi}^2+V(\phi)\right)\, ,\\ \notag &
-2\dot{H}F=\kappa^2\dot{\phi}^2+\ddot{F}_R-H\dot{F}_R
+\frac{4\kappa^2}{3}\rho_r\, ,
\end{align}
\begin{equation}\label{scalareqnofmotion}
\ddot{\phi}+3H\dot{\phi}+V'(\phi)=\frac{Q}{\dot{\phi}}\, ,
\end{equation}
where $F_R=\frac{\partial F}{\partial R}$ and we shall use these
in order to construct an autonomous dynamical system which
controls the dynamics of the $F(R,\phi)$ gravity system. The
autonomous dynamical system of $F(R)-\phi$ gravity system can
easily be obtained if we make use of the following dimensionless
parameters,
\begin{equation}\label{variablesslowdown}
x_1=-\frac{\dot{F}_R(R)}{F_R(R)H},\,\,\,x_2=-\frac{F(R)}{6F_R(R)H^2},\,\,\,x_3=
\frac{R}{6H^2},\,\,\,x_4=\frac{\kappa}{H}\sqrt{\frac{V}{3}},\,\,\,x_5=\frac{\kappa\dot{\phi}}{H},\,\,\,
x_6=\frac{1}{F_R}\, ,
\end{equation}
Recall that the scalar field potential is assumed to be equal to
$V(\phi)=V_0\,e^{-\lambda \phi}$ with $\lambda$ defined in Eq.
(\ref{lambda}), see also Eq. (\ref{potentialapprox}). We shall
express the dynamical variable of the system to be the
$e$-foldings number instead of the cosmic time. So by using the
dimensionless variables (\ref{variablesslowdown}) and the field
equations (\ref{eqnsofmkotion}) and (\ref{scalareqnofmotion}), we
obtain the following autonomous dynamical system,
\begin{align}\label{dynamicalsystemmain}
& \frac{\mathrm{d}x_1}{\mathrm{d}N}=x_1^2-x_1 x_3+3 x_1+2 x_3+
x_5^2x_6-4\, ,
\\ \notag &
\frac{\mathrm{d}x_2}{\mathrm{d}N}=m+x_1 x_2-2 x_2 x_3+4 x_2-4 x_3+8\, ,\\
\notag & \frac{\mathrm{d}x_3}{\mathrm{d}N}=-m-2 x_3^2+8 x_3-8 \, ,
\\ \notag & \frac{\mathrm{d}x_4}{\mathrm{d}N}=-x_3 x_4-\frac{\lambda  x_4
x_5}{2}+2 x_4 \, , \\
\notag & \frac{\mathrm{d}x_5}{\mathrm{d}N}=-x_5+3\lambda x_4^2-x_5x_3 \, , \\
\notag & \frac{\mathrm{d}x_6}{\mathrm{d}N}=x_1 x_6 \, .
\end{align}
with the parameter $m$ being defined to be,
\begin{equation}\label{parameterm}
m=-\frac{\ddot{H}}{H^3}\, .
\end{equation}
Now, the dynamical system of Eq. (\ref{dynamicalsystemmain}) is
not entirely autonomous unless the parameter $m$ is a constant.
This is true for the cases of interest, that is for a quasi-de
Sitter evolution of the form $H(t)=H_0-H_I\,t$ and for a power-law
evolution of the form $H(t)=\frac{\eta}{t}$, with
$\eta=\frac{\beta +2}{3 \beta }$, which describes the power-law
tail inflationary tail that is controlled by the scalar field, if
$\beta<1$. For the quasi-de Sitter evolution, we have that $m=0$
and for the power-law inflationary tail we have
$m=-\frac{2}{\eta^2}$, so in both cases the dynamical system is
autonomous. The effective EoS, defined as,
\begin{equation}\label{weffoneeqn}
w_{eff}=-1-\frac{2\dot{H}}{3H^2}\, ,
\end{equation}
can be written in terms of the parameters of the dynamical system,
and specifically it reads,
\begin{equation}\label{eos1}
w_{eff}=-\frac{1}{3} (2 x_3-1)\, .
\end{equation}
Our aim in this section is to investigate in depth the phase space
of the $F(R,\phi)$ system, revealing any periodicity features and
stable attractors, for the power-law case, in which
$m=-\frac{2}{\eta^2}$ with $\eta=\frac{\beta +2}{3 \beta }$ and
for the $m=0$ case. Let us start with the power-law inflationary
tail case, which interests us the most. Our aim is to reveal
whether the dynamical system reaches the attractor for which the
total EoS is $w_{eff}=\frac{\beta-2}{\beta+2}$. If the system
reaches a stable attractor, which yields
$w_{eff}=\frac{\beta-2}{\beta+2}$ then this would be a proof that
the power-law tail indeed dominates the evolution compared to the
$R^2$ gravity. Unfortunately our results indicate that this is not
true. For $\beta=0.99$ we found unstable and saddle fixed points
for the system, and also none of the fixed points is reached by
the phase space trajectories, which strongly diverge after a few
$e$-foldings. Let us start the analysis by presenting the fixed
points of the system (\ref{dynamicalsystemmain}) for $\beta=0.99$,
which are listed in Table \ref{table0}.
\begin{table}[h!]
  \begin{center}
    \caption{\emph{The Fixed Points of the Dynamical System  of Eq. (\ref{dynamicalsystemmain}) for $\beta=0.99$ and $m=-\frac{2}{\eta^2}$ with $\eta=\frac{\beta +2}{3 \beta }$.}}
    \label{table0}
    \begin{tabular}{|r|r|r|}
     \hline
      \textbf{Name of Fixed Point} & \textbf{Fixed Point Values for $\beta=0.99$}  \\
           \hline
      $P_1^*$ & $(x^*_1,x^*_2,x^*_3,x^*_4x^*_5,x^*_6)=(-2.72291,2.71622,1.00669,0,0,0)$  & Saddle\\ \hline
      $P_2^*$ & $(x^*_1,x^*_2,x^*_3,x^*_4x^*_5,x^*_6)=( 0.729596, -0.736285,1.00669,0, 0,0)$  & Saddle \\ \hline
      $P_3^*$ & $(x^*_1,x^*_2,x^*_3,x^*_4x^*_5,x^*_6)=(-2.72291, 2.71622,1.00669, -0.817861,1.40948, 0)$  & Saddle \\ \hline
      $P_4^*$ & $(x^*_1,x^*_2,x^*_3,x^*_4x^*_5,x^*_6)=(-2.72291,2.71622,1.00669,0.817861,1.40948,0)$
      & Saddle
      \\\hline
      $P_5^*$ & $(x^*_1,x^*_2,x^*_3,x^*_4x^*_5,x^*_6)=(0,-1.00669,1.00669,-0.817861,1.40948, 1)$
      &Saddle
      \\\hline
      $P_6^*$ & $(x^*_1,x^*_2,x^*_3,x^*_4x^*_5,x^*_6)=(0,-1.00669,1.00669,0.817861,1.40948, 1)$
      &Saddle
      \\\hline
      $P_7^*$ & $(x^*_1,x^*_2,x^*_3,x^*_4x^*_5,x^*_6)=(0.729596,-0.736285,1.00669,-0.817861,1.40948, 0)$
      &Saddle
      \\\hline
      $P_8^*$ & $(x^*_1,x^*_2,x^*_3,x^*_4x^*_5,x^*_6)=(0.729596,-0.736285,1.00669,0.817861,1.40948,0)$
      &Saddle
      \\\hline
    \end{tabular}
  \end{center}
\end{table}
All the fixed points of the dynamical system  of Eq.
(\ref{dynamicalsystemmain}) are unstable and share a common
property, $w_{eff}=-0.337793=\frac{\beta-2}{\beta+2}$. So
basically the total EoS is described by a scalar field dominated
evolution, hence the power-law inflationary tail is realized
indeed. However, the dynamics in the phase space indicate that
these unstable fixed points are never reached, regardless the
initial conditions we choose. It seems that, although the fixed
points are indeed unstable, the phase space trajectories never
reach the unstable fixed points and blow-up in the trajectory
space as a function of the $e$-foldings. To show this, in Fig.
\ref{plot1} we present the trajectories in the phase space of the
dynamical system (\ref{dynamicalsystemmain}) for the initial
conditions $x_1(0)=-0.01$, $x_2(0)=0.01$, $x_3(0)=2.05$,
$x_4(0)=0$, $x_5(0)=7$, $x_6(0)=-2$,  and for
$m=-\frac{2}{\eta^2}$, $\beta=0.99$.
\begin{figure}[h]
\centering
\includegraphics[width=35pc]{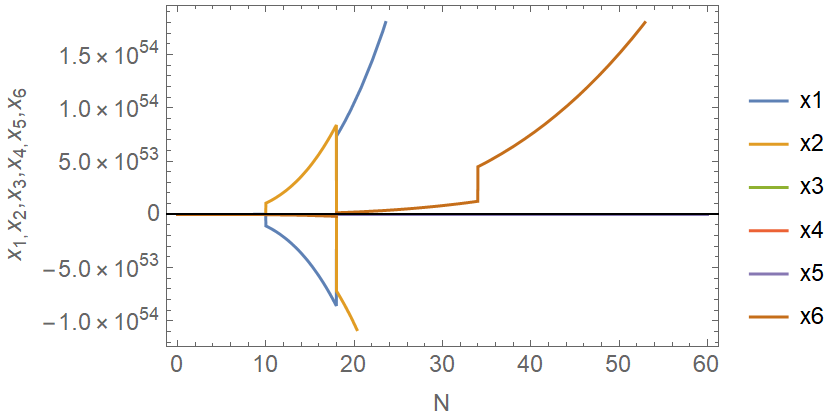}
\caption{{\it{Numerical solutions of the trajectories $x_1(N)$,
$x_2(N)$, $x_3(N)$, $x_4(N)$, $x_5(N)$ and $x_6(N)$ for the
dynamical system (\ref{dynamicalsystemmain}), using the initial
conditions $x_1(0)=-0.01$, $x_2(0)=0.01$, $x_3(0)=2.05$,
$x_4(0)=0$, $x_5(0)=7$, $x_6(0)=-2$,  and for
$m=-\frac{2}{\eta^2}$, $\beta=0.99$.}}} \label{plot1}
\end{figure}
As it can be seen in Fig. \ref{plot1} the trajectories blow-up in
the phase space and the fixed points are never reached. Now this
said behavior occurs for quite a large number of initial condition
sets. Thus the dynamical system is extremely unstable and one
cannot claim easily that the scalar field dominates the evolution.
This would happen for example if initial conditions existed for
which the dynamical system would be attracted on the fixed points
and then would be repelled from these. This does not happen, so
the only scenario for which the TCC issues can be resolved using a
power-law evolution is the one which eliminates the $F(R)$ gravity
once a specific curvature scale is reached before the end of the
$R^2$ inflation. In this way, the $F(R)$ gravity would reduce to a
simple Einstein-Hilbert gravity and the dynamical system
(\ref{dynamicalsystemmain}) would no longer be valid. Such a
scenario will be presented in the next section.

\begin{figure}[h]
\centering
\includegraphics[width=25pc]{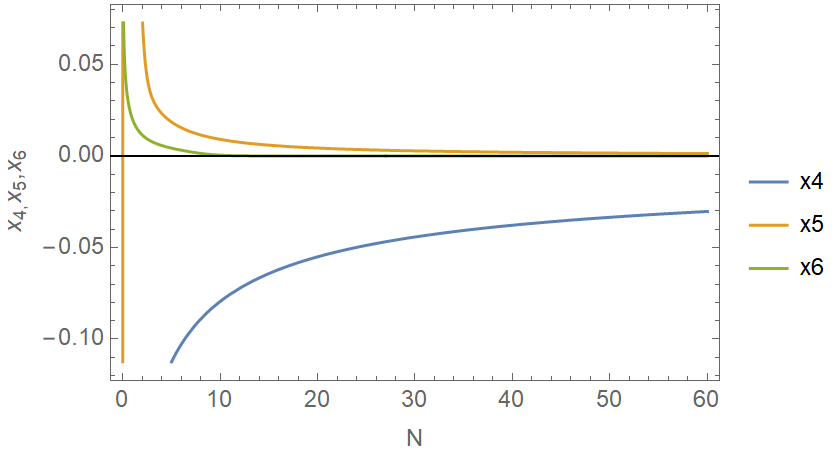}
\includegraphics[width=25pc]{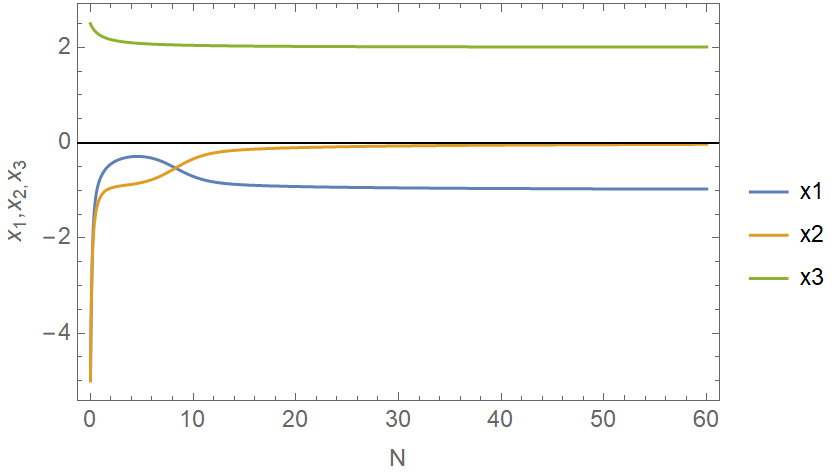}
\caption{{\it{Numerical solutions of the dynamical system
(\ref{dynamicalsystemmain}) $x_1(N)$, $x_2(N)$, $x_3(N)$,
$x_4(N)$, $x_5(N)$ and $x_6(N)$ using the initial conditions
$x_1(0)=-5$, $x_2(0)=-5$, $x_3(0)=2.5$, $x_4(0)=-5$,
$x_5(0)=-0.2$, $x_6(0)=1$,  and for $m=0$ and $\beta=0.99$.}}}
\label{plot2}
\end{figure}

\begin{figure}[h]
\centering
\includegraphics[width=35pc]{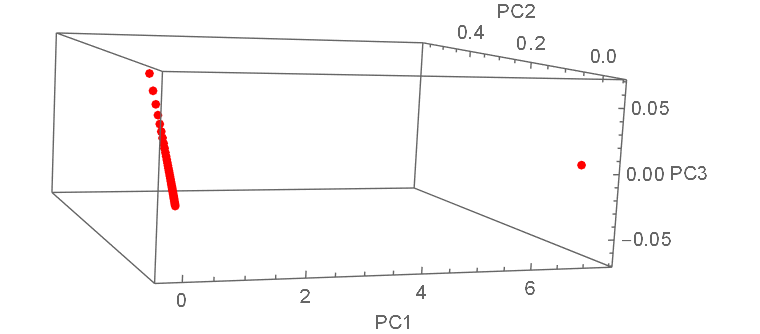}
\caption{PCA analysis of the dynamical system of Eq.
(\ref{dynamicalsystemmain}).} \label{plot3}
\end{figure}

\begin{figure}[h]
\centering
\includegraphics[width=20pc]{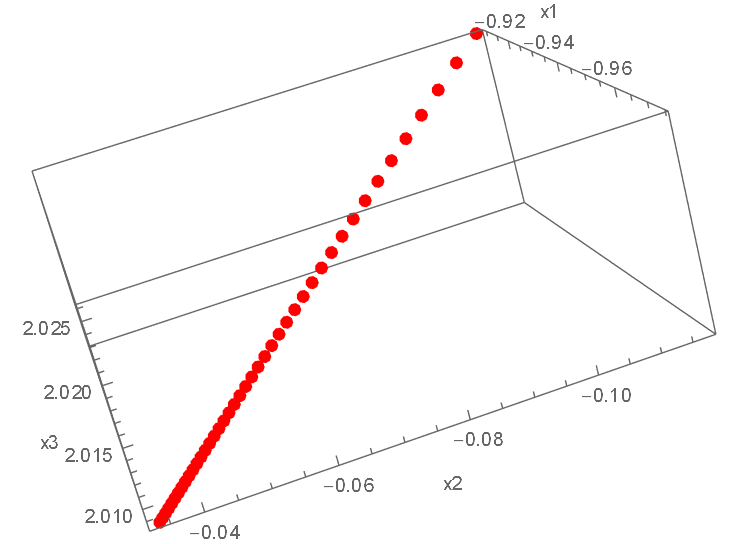}
\includegraphics[width=20pc]{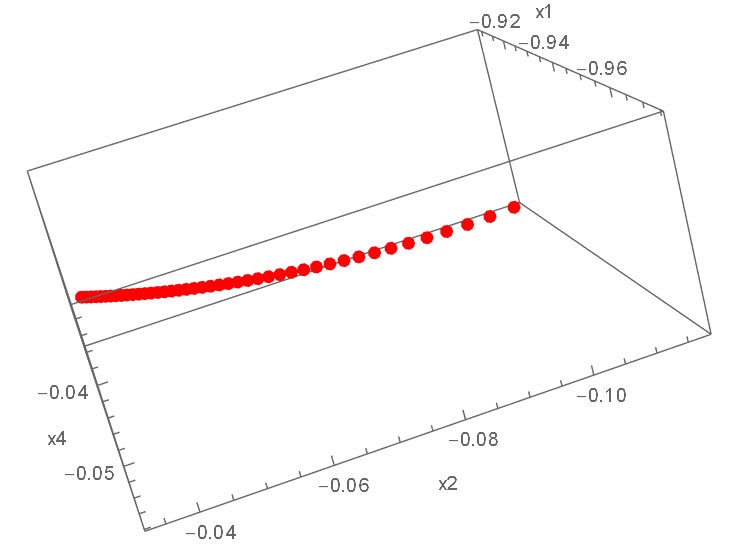}
\includegraphics[width=20pc]{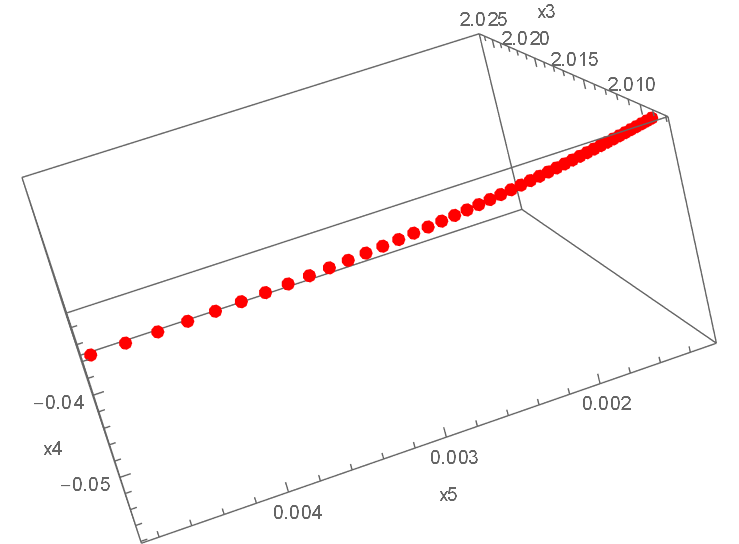}
\caption{Poincare section $x_6=0$ and dynamics for various
variables of the dynamical system of Eq.
(\ref{dynamicalsystemmain}.} \label{plot4}
\end{figure}

\begin{figure}[h]
\centering
\includegraphics[width=20pc]{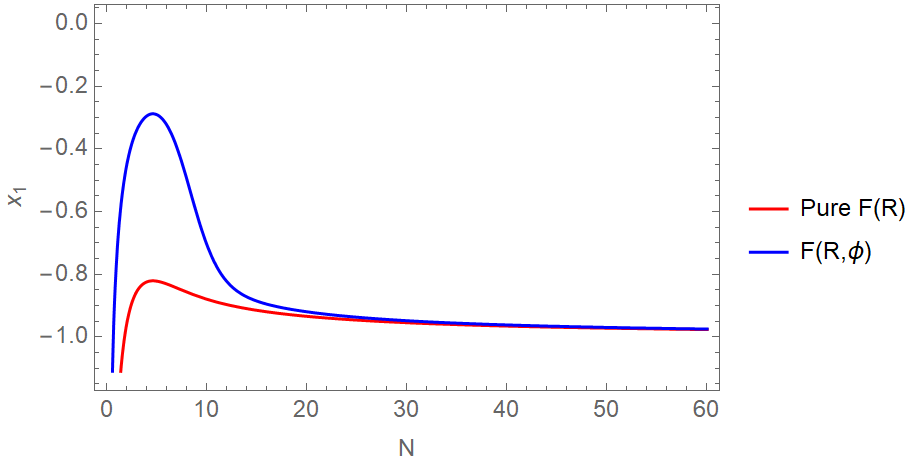}
\includegraphics[width=20pc]{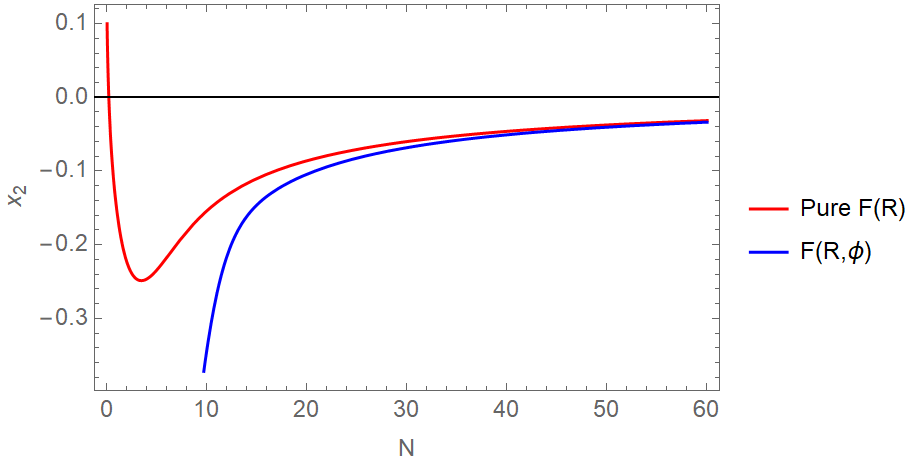}
\includegraphics[width=20pc]{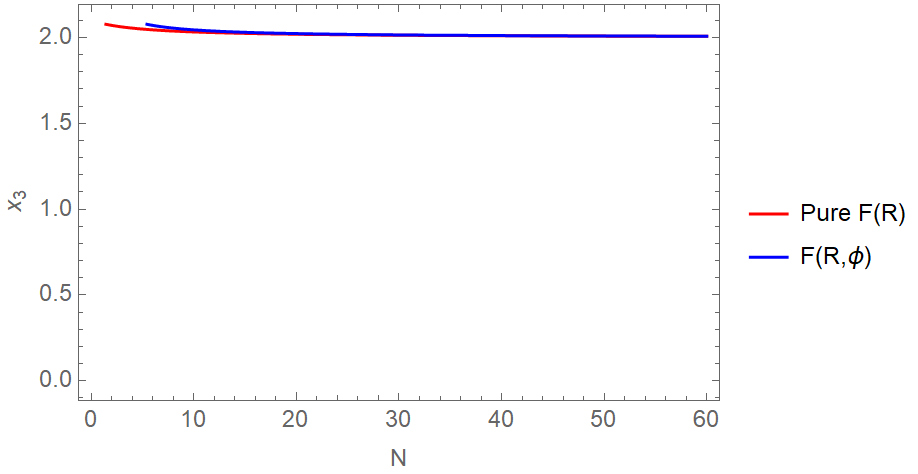}
\caption{Plot of the trajectories for the variables $x_1$, $x_2$
and $x_3$ for the pure $F(R)$ gravity dynamical system
(\ref{dynamicalsystemmainTCC}) and for the combined $F(R,\phi)$
dynamical system (\ref{dynamicalsystemmain}) with the initial
conditions being chosen as $x_1(0)=-5$, $x_2(0)=-5$, $x_3(0)=2.5$,
$x_4(0)=-5$, $x_5(0)=-0.2$, $x_6(0)=1$. We used $m=0$ and
$\beta=0.99$ for the case of the $F(R,\phi)$ dynamical system
(\ref{dynamicalsystemmain}). The red curves correspond to the
trajectories of the pure $F(R)$ gravity dynamical system
(\ref{dynamicalsystemmainTCC}) and the blue curves correspond to
the $F(R,\phi)$ dynamical system (\ref{dynamicalsystemmain}). The
two systems share the same fixed points, because the variables
$x_1$, $x_2$ and $x_3$ converge to the same fixed point in the
phase space.} \label{plot6}
\end{figure}

Now it is worth discussing the de Sitter subspace of the total
phase space of the dynamical system (\ref{dynamicalsystemmain}).
Recall that in this case, $m=0$, thus let us analyze the de Sitter
scenario in order to reveal how the phase space behaves in this
case. We start with the fixed points of the system, with
$\beta=0.99$ and these are presented in Table \ref{table1}.
\begin{table}[h!]
  \begin{center}
    \caption{\emph{The Fixed Points of the Dynamical System  of Eq. (\ref{dynamicalsystemmain}) for $\beta=0.99$ and $m=0$.}}
    \label{table1}
    \begin{tabular}{|r|r|r|}
     \hline
      \textbf{Name of Fixed Point} & \textbf{Fixed Point Values for $\beta=0.99$} & \textbf{Stability} \\
           \hline
      $P_1^*$ & $(x^*_1,x^*_2,x^*_3,x^*_4x^*_5,x^*_6)=(0,x_2,2,0,0,x_6)$  & Unstable\\ \hline
      $P_2^*$ & $(x^*_1,x^*_2,x^*_3,x^*_4x^*_5,x^*_6)=(0,x_2,2,0,0,0)$  & Unstable \\ \hline
      $P_3^*$ & $(x^*_1,x^*_2,x^*_3,x^*_4x^*_5,x^*_6)=(-1, 0,2,0,0, 0)$  & Unstable
      \\\hline
    \end{tabular}
  \end{center}
\end{table}
As it can be seen in Table \ref{table1}, the dynamical system of
Eq. (\ref{dynamicalsystemmain}) for the de-Sitter subspace has
three unstable fixed points. The phase space structure is quite
interesting in this case, since it has some lower dimensional
stability as we show shortly, but also has an inherent
consistency, compared with the previous power-law case, because
the trajectories in the phase space tend to the fixed point
$P_3^*$. We can clearly show this by solving numerically the
dynamical system (\ref{dynamicalsystemmain}) using various initial
conditions. Our results are presented in Fig. \ref{plot2}, using
the initial conditions $x_1(0)=-5$, $x_2(0)=-5$, $x_3(0)=2.5$,
$x_4(0)=-5$, $x_5(0)=-0.2$, $x_6(0)=1$, and for $m=0$ and
$\beta=0.99$.
As we can see in Fig. \ref{plot2} all the solutions tend to the
fixed point  $P_3^*$, however the parameter $x_6$ does not reach
the fixed point value $x_6=0$ even after 60 $e$-foldings. This
feature is true for a large number of sets of initial conditions.
The behavior of the trajectories though shows some formal stable
fixed point dynamical evolution, because eventually the fixed
point $P_3^*$ is reached by most of the variables. Also the total
EoS is $w_{eff}=-1$ due to the fact that $x_3=2$, recall Eq.
(\ref{eos1}) which relates the total EoS with the parameter $x_3$.
It is worth studying more the de Sitter subspace of the total
phase space, so let us consider some structures that can be found
in the phase space for $m=0$. Let  us start with the Principal
Component Analysis (PCA) which can reveal the actual dynamics of
the dynamical system of Eq. (\ref{dynamicalsystemmain}) and reduce
the dimensionality of the dynamical system, using only the fast
variables that determine truly the dynamics of the system. If the
dynamical system of Eq. (\ref{dynamicalsystemmain}) has intrinsic
lower-dimensional structure, the PCA projection can in principle
reveal it. If the first three components in the PCA projection
have or generate some characteristic structure or pattern, then
the six-dimensional initial system of Eq.
(\ref{dynamicalsystemmain}) effectively behaves like a three-
dimensional one. If the projection does not preserve any key
features whatsoever, more dimensions may be needed for the
analysis. We shall use the PCA analysis in order to reveal the
dominant modes of the dynamical system of Eq.
(\ref{dynamicalsystemmain}) and the results of our PCA analysis
can be found in Fig. \ref{plot3}. Apparently, the resulting
pattern indicates the presence of multiple isolated stable fixed
points, with no signs of periodicity or chaos. This is somewhat
exciting due to the presence of the unstable fixed points we
presented earlier for the dynamical system of Eq.
(\ref{dynamicalsystemmain}). The lower-dimensional subspace of the
total phase space seems stable though, thus there must be an
inherent structure in the dynamical system.
Now it is worth examining the Poincare sections of the dynamical
system of Eq. (\ref{dynamicalsystemmain}). The Poincare sections
reduce the complexity of the dynamical system and indicate which
variables drive the dynamical evolution. In Figs. \ref{plot4} and
we present the Poincare sections corresponding to $x_6=0$. As it
can be seen, there exist various isolated points of equilibrium
that capture the dynamics and it seems that the variable $x_6$
does not affect significantly the dynamics of the system. Both the
Poincare sections and the PCA analysis indicate the existence of
isolated stable points for various initial conditions, however no
periodicity or chaos is revealed in the dynamical system of Eq.
(\ref{dynamicalsystemmain}).
Now the intriguing question is whether $F(R)$ gravity dominates
the evolution over the scalar field with exponential potential.
Let us get into some details on this and we start with the EoS of
the scalar field which is defined as,
\begin{equation}\label{eosscalardefinition1}
w_{\phi}=\frac{\frac{\dot{\phi}^2}{2}-V}{\frac{\dot{\phi}^2}{2}+V}\,
,
\end{equation}
and for the potential chosen as in Eq. (\ref{potentialapprox}) it
should result to Eq. (\ref{eosscalarfinal}) and for $\beta=0.99$
should take the value $w_{\phi}=-0.337793$. We can express the
scalar field EoS (\ref{eosscalardefinition1}) in terms of the
variables $x_4$ and $x_5$ defined in Eq. (\ref{variablesslowdown})
and we have,
\begin{equation}\label{newdefscalareos}
w_{\phi}=\frac{\frac{x_5^2}{2}-3x_4^2}{\frac{x_5^2}{2}+3x_4^2}\, .
\end{equation}
We can plot the scalar field EoS of Eq. (\ref{newdefscalareos}) as
a function of the $e$-foldings and this done in Fig. \ref{plot4}.
\begin{figure}[h]
\centering
\includegraphics[width=30pc]{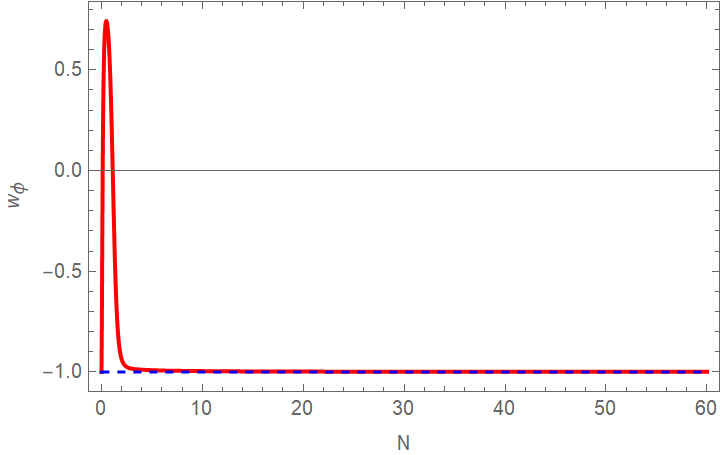}
\caption{The scalar field EoS $w_{\phi}$ as a function of the
$e$-foldings number, for the dynamical system
(\ref{dynamicalsystemmain}) with initial conditions $x_1(0)=-5$,
$x_2(0)=-5$, $x_3(0)=2.5$, $x_4(0)=-5$, $x_5(0)=-0.2$, $x_6(0)=1$,
and for $m=0$ and $\beta=0.99$.} \label{plot5}
\end{figure}
As it can be seen in Fig. \ref{plot5} the behavior of the scalar
field EoS is intriguing because after a few $e$-foldings it
settles in the value $w_{\phi}\simeq -1$, which contradicts the
expected value $w_{\phi}=-0.337793$ that it should have. Now it is
worth recalling the pure $F(R)$ gravity phase space in order to
completely understand what is going on in this obscure situation.
The analysis of the pure $F(R)$ gravity phase space was performed
in Ref. \cite{Odintsov:2017tbc} so at this point we shall recall
the essential features and results of that analysis. When the
$F(R)$ gravity phase space is considered solely, the following
variables are used,
\begin{equation}\label{variablesslowdownTCC}
x_1=-\frac{\dot{F_R}(R)}{F_R(R)H},\,\,\,x_2=-\frac{F(R)}{6F_R(R)H^2},\,\,\,x_3=
\frac{R}{6H^2}\, ,
\end{equation}
and the pure $F(R)$ gravity dynamical system becomes,
\begin{align}\label{dynamicalsystemmainTCC}
& \frac{\mathrm{d}x_1}{\mathrm{d}N}=-4-3x_1+2x_3-x_1x_3+x_1^2\, ,
\\ \notag &
\frac{\mathrm{d}x_2}{\mathrm{d}N}=8+m-4x_3+x_2x_1-2x_2x_3+4x_2 \, ,\\
\notag & \frac{\mathrm{d}x_3}{\mathrm{d}N}=-8-m+8x_3-2x_3^2 \, ,
\end{align}
where $m$ is again,
\begin{equation}\label{parametermTCC}
m=-\frac{\ddot{H}}{H^3}\, .
\end{equation}
The dynamical system of Eq. (\ref{dynamicalsystemmainTCC}) has the
following fixed points,
\begin{equation}\label{fixedpointdesitterTCC}
\phi_*^1=(-1,0,2),\,\,\,\phi_*^2=(0,-1,2)\, ,
\end{equation}
with $\phi_*^1$ being the stable fixed point and the other fixed
point $\phi_*^2$ is the unstable one with both the fixed points
being de Sitter fixed points, since , since for both $x_3=2$ and
hence, the total EoS is $w_{eff}=-1$. Now let us compare the
trajectories of the pure $F(R)$ gravity and the $F(R,\phi)$
gravity phase space. This is done in Fig. \ref{plot6}, where we
plot the trajectories for the variables $x_1$, $x_2$ and $x_3$ for
the pure $F(R)$ gravity dynamical system
(\ref{dynamicalsystemmainTCC}) and for the $F(R,\phi)$ dynamical
system (\ref{dynamicalsystemmain}) with initial conditions
$x_1(0)=-5$, $x_2(0)=-5$, $x_3(0)=2.5$, $x_4(0)=-5$,
$x_5(0)=-0.2$, $x_6(0)=1$, and for $m=0$ and $\beta=0.99$. With
red curves we plot the trajectories of the pure $F(R)$ gravity
dynamical system (\ref{dynamicalsystemmainTCC}) and with blue
curves the $F(R,\phi)$ dynamical system
(\ref{dynamicalsystemmain}). As it can be seen in Fig.
\ref{plot6}, the two systems share the same fixed points, since
the variables $x_1$, $x_2$ and $x_3$ have the same behavior after
some $e$-foldings. However, we need to note that the fixed point
is unstable in the combined $F(R,\phi)$ dynamical system
(\ref{dynamicalsystemmain}) while it is stable in the pure $F(R)$
gravity dynamical system (\ref{dynamicalsystemmainTCC}).
By looking Fig. \ref{plot6} one thing is certain, it seems that
the $F(R)$ gravity controls the dynamics if the evolution is a de
Sitter one, and as we saw earlier, for the power-law tail
evolution, the dynamical system is strongly unstable with the
trajectories blowing up and never reaching the existing unstable
fixed points. Thus, in order for the power-law inflationary tail
to solve the TCC problems of standard inflation, one needs to
eliminate effectively the non-trivial $F(R)$ part from the field
equations, thus leaving a pure scalar field dominated evolution.
We present an effective theory of this sort in the next section.

\section{A Way to Satisfy the TCC Requirements: Choice of a Suitable $F(R)$ Gravity that Ensures an Exact Scalar Field Driven Power-law Inflationary Tail}

\begin{figure}[h]
\centering
\includegraphics[width=30pc]{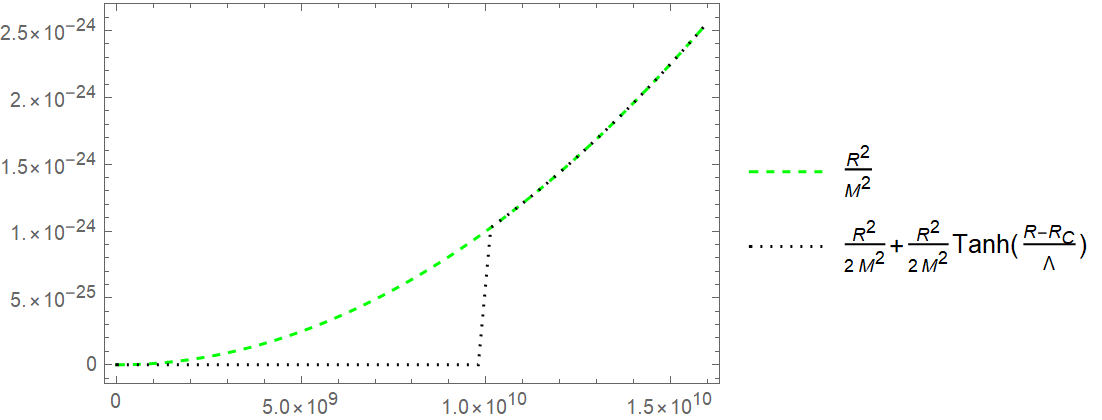}
\caption{The comparison of the $\frac{R^2}{2M^2}+\tanh \left(
\frac{R-R_c}{\Lambda}\right)\,\frac{R^2}{2M^2}$ and
$\frac{R^2}{M^2}$ terms for $R_c\sim 10^{10}\,$eV$^2$. As it can
be seen, when the curvature lowers below the critical value $R_c$,
the $\frac{R^2}{2M^2}+\tanh \left(
\frac{R-R_c}{\Lambda}\right)\,\frac{R^2}{2M^2}$ term is
effectively zero.} \label{plot7}
\end{figure}

\begin{figure}[h!]
\centering
\includegraphics[width=35pc]{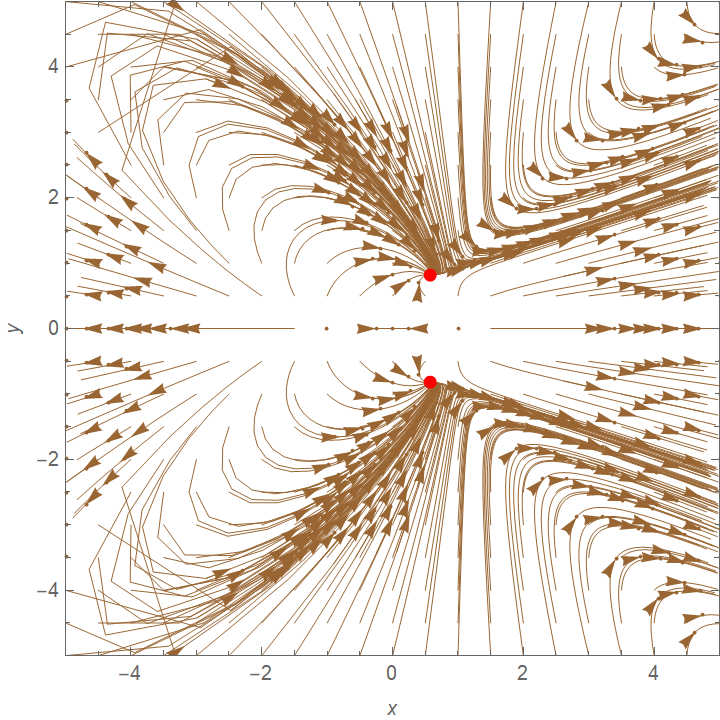}
\caption{Phase space trajectories for the dynamical system
(\ref{dynamicalsystemtwodimensionaldynsubsystemain}) for various
initial conditions. The red dots represent the fixed points
$P_3^*=(x_*,y_*)=(0.575416,-0.817861)$ and
$P_4^*=(x_*,y_*)=(0.575416,0.817861)$.} \label{plot8}
\end{figure}

As we demonstrated in the previous sections, in order for the
power-law tail to control the dynamics of inflation for the last
$e$-foldings of the inflationary era, one must clearly cut off the
non-linear $F(R)$ gravity part. In this section we shall use a
phenomenological $F(R)$ gravity model which can actually allow the
power-law tail to be realized. The model we shall use is a
exponential deformation of the $R^2$ model, of the form,
\begin{equation}\label{FRGravitydeformationmain}
F(R)=R+\frac{R^2}{2M^2}+\tanh \left(
\frac{R-R_c}{\Lambda}\right)\,\frac{R^2}{2M^2}+f_{DE}(\frac{R}{\Lambda})\,
.
\end{equation}
These models are highly motivated by arguments related to the
unified description of inflation with the dark energy era, as was
demonstrated in Ref. \cite{modelagnosticFR}. The term
$f_{DE}(\frac{R}{\Lambda})$ can be chosen on the phenomenological
basis of Ref. \cite{modelagnosticFR} and can be a power-law term
of the Ricci scalar, or some exponential model. This term however
cannot dominate over the scalar field at late times, due to the
fact that the scalar field potential parameter $V_0$ in Eq.
(\ref{potentialapprox}) is constrained by the Planck data
\cite{Planck:2018jri} to be $V_0\simeq 9\times 10^{-11}M_p^4$ thus
the scalar potential would certainly dominate over a wide range of
dark energy driving $F(R)$ gravity terms. So let us focus on the
inflationary era, in which case the $f_{DE}(\frac{R}{\Lambda})$
part is strongly subdominant and both the scalar field and $R^2$
terms dominate over it. Now let us demonstrate how the $F(R)$
gravity behaves during the inflationary era. The critical scalar
curvature $R_c$ in Eq. (\ref{FRGravitydeformationmain})
corresponds to the value of the curvature scalar when the
power-law tail should take over the control of the dynamics of the
cosmological evolution, so it corresponds to some curvature near
the end of the slow-roll era. The behavior of the term $\tanh
\left( \frac{R-R_c}{\Lambda}\right)$ during the inflationary era
is,
\begin{equation}\label{interactionterm2}
\tanh \left( \frac{R-R_c}{\Lambda} \right)=\left\{
\begin{array}{c}
  1,\,\,\,\mathrm{when}\,\,\, R> R_c \\
  0,\,\,\,\mathrm{when}\,\,\, R\sim R_c  \\
  -1,\,\,\,\mathrm{when}\,\,\, R< R_c \, ,
\end{array}\right.
\end{equation}
Thus the $F(R)$ gravity function of Eq.
(\ref{FRGravitydeformationmain}) behaves in the following way
during the inflationary era,
\begin{equation}\label{interactionterm2}
F(R)=\left\{
\begin{array}{c}
  R+\frac{R^2}{M^2},\,\,\,\mathrm{when}\,\,\, R> R_c \\
  R+\frac{R^2}{2M^2},\,\,\,\mathrm{when}\,\,\, R\sim R_c  \\
  R,\,\,\,\mathrm{when}\,\,\, R< R_c \, ,
\end{array}\right.
\end{equation}
Hence, with the phenomenological choice of Eq.
(\ref{FRGravitydeformationmain}) the $R^2$ part of the $F(R)$
gravity is effectively switched off during the last stages of
inflation. At the beginning of the inflationary era, the $F(R)$
gravity controls the dynamics with the dominant $R^2$ term driving
the evolution, then once the curvature lowers down to the critical
value $R_c$ the $F(R)$ gravity is basically described by an
Einstein-Hilbert term. We have shown this in Fig. \ref{plot7}
where we present the behavior of the $F(R)$ gravity during the
inflationary era, taking $R_c\sim 10^{10}\,$eV$^2$ as an example,
and using the phenomenological values of $M$ and $\Lambda$. As it
can be seen in Fig. \ref{plot7}, when the curvature lowers below
the critical curvature, $R_c\sim 10^{10}\,$eV$^2$, the $R^2$ term
is switched off and the effective $F(R)$ gravity is described by
the Einstein-Hilbert term solely.
Thus with the choice of Eq. (\ref{FRGravitydeformationmain}), the
scalar field basically dominates the evolution for values of the
curvature lower the critical curvature $R_c$, and therefore, the
Friedmann equation becomes,
\begin{equation}\label{subsystemfriedmann}
3H^2=\frac{\kappa^2\dot{\phi}^2}{2}+V\, .
\end{equation}
Also the total EoS parameter $w_{tot}$ for the scalar field
dominated inflationary epoch is,
\begin{equation}\label{totaleosparameter}
w_{tot}=\frac{P_{\phi}}{\rho_{\phi}}=w_{\phi}\Omega_{\phi}\, ,
\end{equation}
Using the dimensionless variables,
\begin{equation}\label{variablesdynamicalsystem}
x^2=\frac{\kappa^2\dot{\phi}^2}{6H^2},\,\,\,y^2=\frac{\kappa^2V}{3H^2}\,
,
\end{equation}
we have for the scalar EoS parameter and the energy density,
\begin{equation}\label{constraintsnew}
w_{\phi}=\frac{x^2-y^2}{x^2+y^2},\,\,\,\Omega_{\phi}=x^2+y^2= 1\,
,
\end{equation}
and we can construct the following autonomous dynamical system for
the scalar field \cite{Boehmer:2008av},
\begin{align}\label{dynamicalsystemtwodimensionaldynsubsystemain}
& \frac{\mathrm{d}x}{\mathrm{d}N}=-3x+\frac{\lambda \sqrt{6}}{2}y^2+\frac{3 x}{2}\left(1+x^2-y^2 \right)\, ,              \\
\notag & \frac{\mathrm{d}y}{\mathrm{d}N}=-\frac{\lambda
\sqrt{6}}{2}x\,y+\frac{3 y}{2}\left(1+x^2-y^2 \right)\, ,
\end{align}
using again the $e$-foldings number $N$ as the dynamical variable.
For $\beta=0.99$ the fixed points of the system, their stability,
and their physical significance are presented in Table
\ref{tablescalaronly}. As it can be seen, the only physically
acceptable fixed points are $P_1^*$, $P_3^*$ and $P_4^*$ and from
these the only stable are $P_3^*$ and $P_4^*$ which perfectly
describe the desired EoS parameter value $w_{tot}=-0.337793$ which
corresponds to the scalar field for $\beta=0.99$.
\begin{table}[h!]
  \begin{center}
    \caption{\emph{Phase Space Fixed Points, Stability and Physical Significance of the Dynamical System (\ref{dynamicalsystemtwodimensionaldynsubsystemain}) for $\beta=0.99$.}}
    \label{tablescalaronly}
    \begin{tabular}{|r|r|r|r|r|}
     \hline
      \textbf{Name of Fixed Point} & \textbf{Fixed Point Values for $\beta=0.99$} & Stability & $\Omega_{\phi}$ & $w_{\phi}$ \\
           \hline
      $P_1^*$ & $(x_*,y_*)=(-1,0)$ & unstable & 1 & 1  \\\hline
      $P_2^*$ & $(x_*,y_*)=(0,0)$  & unstable & &  \\\hline
      $P_3^*$ & $(x_*,y_*)=(0.575416,-0.817861)$  & stable & 1 & -0.337793  \\\hline
      $P_4^*$ & $(x_*,y_*)=(0.575416,0.817861)$  & stable & 1 & -0.337793  \\\hline
      $P_5^*$ & $(x_*,y_*)=(0.868936,0.868936)$ & unstable & 1.5101 & 0  \\\hline
      $P_6^*$ & $(x_*,y_*)=(0.868936,-0.868936)$  & unstable & 1.5101 & 0 \\\hline
    \end{tabular}
  \end{center}
\end{table}
Also, in Fig. \ref{plot8} we present the phase space trajectories
of the dynamical system
(\ref{dynamicalsystemtwodimensionaldynsubsystemain}), including
with red dots the fixed points
$P_3^*=(x_*,y_*)=(0.575416,-0.817861)$ and
$P_4^*=(x_*,y_*)=(0.575416,0.817861)$ as red dots. As it can be
seen in Fig. \ref{plot8}, all the trajectories in the phase space
of the dynamical system
(\ref{dynamicalsystemtwodimensionaldynsubsystemain}) are attracted
to the stable fixed points $P_3^*$ and $P_4^*$. Thus, in this
section we showed that if the non-linear part of the $F(R)$
gravity is switched off near the end of the slow-roll era, the
dynamics of the cosmological system is clearly dominated by the
scalar field, driving the system to the fixed points
$P_3^*=(x_*,y_*)=(0.575416,-0.817861)$ and
$P_4^*=(x_*,y_*)=(0.575416,0.817861)$ which yield an EoS parameter
$w_{tot}=-0.337793$.
Now as a final comment it would be interesting to consider the
late-time dynamics of the resulting quintessence-like theory which
is controlled by the scalar field. The scalar field potential
would dominate over a wide range of dark energy $F(R)$ gravity
models as we mentioned at the beginning of this section, if the
scalar field values are close to the Planck scale $M_p$. This
study however exceeds the purposes of this work and we leave it
for future work.

\section{The Graceful Exit from Inflation Issue with Power-law Inflation}

The power-law inflationary regime yields a constant first
slow-roll index, thus it is not possible to end inflation in this
scenario, if the scalar field $\phi$ that drives the power-law era
is considered by itself without the presence of other fields.
However, the scalar field $\phi$ is a remnant of some ultraviolet
completion of the standard model, thus it may have couplings to
other scalar fields and also to fermions. These couplings may
actually end the inflationary regime controlled by the scalar
field. In this section we shall present two mechanisms that may
end the inflationary era controlled by the scalar field, the first
mechanism employs an auxiliary scalar field $\chi$, a so-called
waterfall field, thus giving rise to some sort of hybrid inflation
\cite{Felder:2001kt,Linde:1993cn,Copeland:1994vg,Copeland:2000hn,Lee:2023dcy},
and the second employs the coupling of the scalar field $\phi$ to
a fermion \cite{Coleman:1973jx,Bostan:2019fvk}. Let us analyze in
detail these two mechanisms.

Before starting, it is important to discuss why ending the
power-law inflationary era is important. Specifically, why the
graceful exit from inflation is important, and how this will help
resolving completely the Trans-Planckian issues of inflation. The
exit from inflation is important. So after the power-law
inflationary era is realized, the Hubble horizon still shrinks,
but in a slower way compared with the slow-roll era. Hence the
Hubble horizon shrinks slowly, thus if inflation lasted for ever,
the Trans-Planckian modes would exit the horizon eventually. Thus
inflation must end, this is why we proposed two mechanisms that
may end the power-law inflation era, and thus provide a complete
resolution for the Trans-Planckian issue.

\subsection{Hybrid Inflation with a Waterfall Field}

For the Hybrid inflation we shall consider the scalar field $\phi$
and its coupling to a waterfall-type auxiliary scalar field
$\chi$. The scalar field $\phi$ will play the role of the inflaton
and slowly rolls in a power-law way and drives inflation, while
the waterfall field $\chi$ will end the inflationary regime of the
field $\phi$ through a symmetry-breaking instability caused by the
effective potential of the two scalars. The proposed potential
could have the following form, the potential is given by:
\begin{equation}\label{hybrid}
V(\phi, \chi) = V_0 e^{-\lambda \phi / M_{p}} + \frac{1}{2} g^2
\phi^2 \chi^2 + \frac{\lambda_{\chi}}{4} \left( \chi^2 - v^2
\right)^2,
\end{equation}
where $g$ is the coupling constant between the scalar field $\phi$
and the waterfall field $\chi$. The interaction coupling
$\lambda_{\chi}$ and also $v$ define the symmetry-breaking sector
of the waterfall field $\chi$. During the power-law inflation, the
field $\chi$ is stabilized at $\chi = 0$ due to its large positive
effective mass for sufficiently large $\phi$. Hence, the
inflationary trajectory follows the line $\chi=0$ and the
effective mass of the waterfall field is,
\begin{equation} m_{\chi}^2(\phi)=\frac{\partial^2 V(\phi, \chi)}{\partial \chi^2}\Big{|}_{\chi=0} = -\lambda_{\chi} v^2 + g^2 \phi^2
\, ,
\end{equation}
and the critical inflaton value $\phi_c$ at which the effective
mass of the waterfall field is zero is equal to,
\begin{equation} m_{\chi}^2(\phi_c) = 0 \quad \Rightarrow \quad
\phi_c = \frac{\sqrt{\lambda_{\chi}}}{g} v.
\end{equation}
Then we have the following symmetry breaking pattern in the
$\chi$-sector: For $\phi > \phi_c$, we have $m_{\chi}^2> 0$ and
$\chi = 0$, thus the waterfall scalar is stable around the origin
in the $\chi$ direction of the effective potential. For $\phi <
\phi_c$, we have $m_{\chi}^2 < 0$ and therefore the $\chi = 0$
minimum becomes unstable, thus this instability triggers a rapid
symmetry-breaking transition. Hence in this mechanism, inflation
ends when the scalar field $\phi$ reaches the critical value
$\phi_c$ at which point, the effective mass squared of the
waterfall field $\chi$ becomes negative. The field $\chi$ rapidly
rolls to its true vacuum situated at the value $\chi = \pm v$,
thus ending the inflationary regime of the scalar field $\phi$ via
a second-order (or sometimes first-order) phase transition, hence
the name waterfall scalar field \cite{Felder:2001kt,Lee:2023dcy}.
In the context of this mechanism one can even have a grasp on the
total duration of the power-law regime, because inflation ends on
the critical value  of the scalar field $\phi_c$ which depends on
the symmetry breaking in the $\chi$ sector. Note that in this
case, the free couplings of the inflation $\phi$ to the waterfall
scalar does not affect the power-law inflationary phase, it
affects only though the ending of the power-law phase which can be
controlled by $\phi_c = \frac{\sqrt{\lambda_{\chi}}}{g} v$. Thus
the model is open for rich phenomenology in the post-power-law
inflationary phase.

\subsection{Ending Power-law Inflation with Fermion Loops}

\begin{figure}[h!]
\centering
\includegraphics[width=25pc]{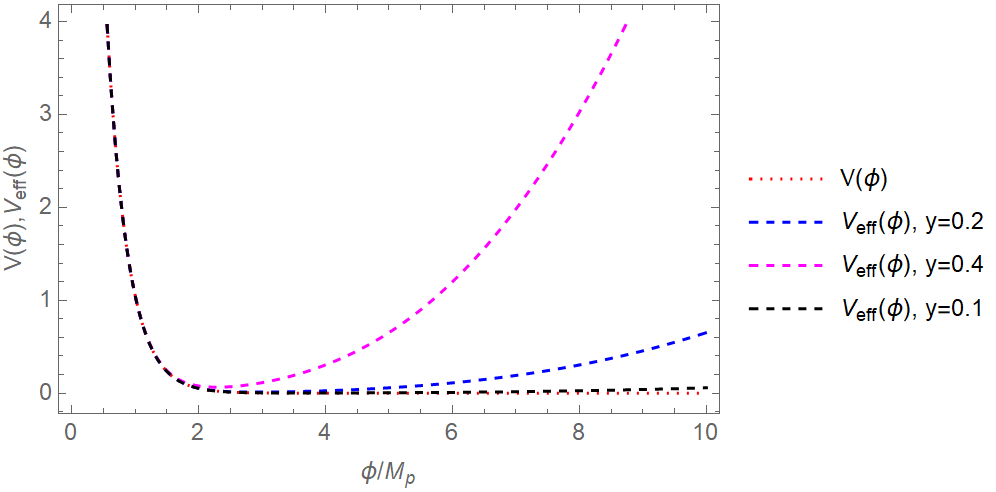}
\caption{The effective potential $V_{eff}(\phi)$ versus the
ordinary potential $V(\phi)$ for various values of the Yukawa
coupling $y$. We used $y=0.1$ (black curve), $y=0.4$ (blue curve)
and $y=0.9$ (magenta curve), while we chose the renormalization
scale to be $\mu=10\,M_p$.} \label{extra}
\end{figure}

\begin{figure}[h!]
\centering
\includegraphics[width=25pc]{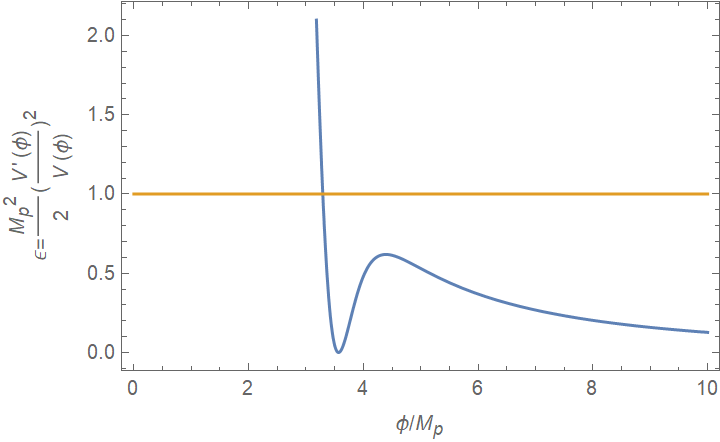}
\caption{The first slow-roll index,
$\epsilon=\frac{M_p^2}{2}\left(\frac{V'(\phi)}{V(\phi)} \right)^2$
for the effective potential $V_{eff}(\phi)$ for $y=0.1$ and
$\mu=10\,M_p$. As it can be seen, the first slow-roll index
crosses the $\epsilon=1$ line and thus the power-law inflationary
era is possible to end.} \label{extra1}
\end{figure}

Another way to terminate the power-law inflation regime of the
scalar field $\phi$ is to add interactions of the scalar field
with fermions, and consider quantum corrections arising from the
fermionic interactions which can modify the inflaton potential and
thus  provide a graceful exit to the power-law regime
\cite{Coleman:1973jx,Bostan:2019fvk}. Let us introduce a Yukawa
interaction between the scalar field $\phi$ and a Dirac fermion
$\psi$, of the form
\begin{equation} \mathcal{L}_{\text{int}} = - y \phi \bar{\psi}
\psi,
\end{equation}
where $y$ is the Yukawa coupling. This Yukawa interaction gives
the following effective mass to the Dirac fermion $m_{\psi}(\phi)
= y \phi$. At the quantum level, the Yukawa interaction modifies
the scalar field potential through one-loop radiative corrections,
and specifically, through the Coleman-Weinberg one-loop effective
potential. This Dirac fermion one-loop correction to the scalar
field effective potential has the following form
\cite{Coleman:1973jx,Bostan:2019fvk},
\begin{equation} \Delta V_{\text{1-loop}}(\phi) = -
\frac{N_f}{8\pi^2} m_{\psi}^4(\phi) \ln \left(
\frac{m_{\psi}^2(\phi)}{\mu^2} \right),
\end{equation}
where $N_f$ is the number of fermionic degrees of freedom and
$\mu$ is the renormalization scale. Substituting $m_{\psi}(\phi) =
y \phi$, the total scalar field effective potential becomes
\begin{equation}\label{effectivepot}
V_{\text{eff}}(\phi) = V_0 e^{-\lambda \phi / M_{p}} - \frac{N_f
y^4 \phi^4}{8\pi^2} \ln \left( \frac{y^2 \phi^2}{\mu^2} \right)\,
,
\end{equation}
therefore, the first slow-roll index is no longer constant, and
thus inflation can come to an end. Both the fermion loop and the
waterfall mechanism have their inherent appeal, however the
waterfall mechanism offers more possibilities for model building.
Also it is possible to end inflation by using particle creation,
although this type of scenarios is used in warm inflation
frameworks. Finally, we need to note that in the fermion loop
case, the Yukawa coupling might affect the duration of the
inflationary phase, in order to have a significant duration for
the power-law case, one must choose the Yukawa coupling
accordingly. For example we plotted the effective potential
$V_{eff}(\phi)$ versus the ordinary potential $V(\phi)$ for
various values of the Yukawa coupling $y$ in Fig. \ref{extra} and
it can clearly be seen how the Yukawa coupling affects the form of
the potential.
As it can be seen in Fig. \ref{extra}, the exponential potential
is similar for all the cases for small values of the scalar field,
sub-Planck ones specifically, but the potential deviates
significantly for large values of the Yukawa coupling for
trans-Planckian values of the scalar field, when the Yukawa
coupling is larger than $y\sim 0.1$. Hence, one must choose small
values of the Yukawa coupling, if this mechanism is to function
properly, without significantly affecting the power-law
inflationary regime. However, the presence of the logarithm always
affects the final form of the potential and even terminates the
power-law era, which is the important feature of these radiative
fermion loops. Specifically, the most optimal scenario for our
analysis, based on Fig. \ref{extra} would be the case with $y=0.1$
or even smaller Yukawa coupling. The scenario does not deviate
from the power-law evolution, and also it is also possible to
terminate the power-law inflationary era. To see this, let us
recall that the first slow-roll index has the form,
$\epsilon=\frac{M_p^2}{2}\left(\frac{V'(\phi)}{V(\phi)}
\right)^2$, so for the effective potential (\ref{effectivepot})
the first slow-roll index takes the form,
\begin{equation}\label{firstslowrollpowerlaw}
\epsilon=\frac{2 M_p^2 \left(4 \pi ^2 \lambda  V_0+2 y^4 \phi ^3
e^{\lambda  \phi } \log \left(\frac{y^2 \phi ^2}{\mu
^2}\right)+y^4 \phi ^3 e^{\lambda  \phi }\right)^2}{\left(y^4 \phi
^4 e^{\lambda  \phi } \log \left(\frac{y^2 \phi ^2}{\mu
^2}\right)-8 \pi ^2 V_0\right)^2}\, ,
\end{equation}
so solving the equation $\epsilon(\phi)=1$ might be impossible in
order to see whether inflation terminates. We can do a numerical
plot to reveal this, so in Fig. \ref{extra1} we plot the behavior
of the first slow-roll index (\ref{firstslowrollpowerlaw}) as a
function of the scalar field for $\mu=10\,M_p$ and $y=0.1$.
As it can be seen in Fig. \ref{extra1}, the first slow-roll index
$\epsilon$ has the desirable behavior since it is smaller than
unity for larger values of the scalar field, and as the scalar
field values drop, it becomes unity (it crosses the $\epsilon=1$
orange line) for some model-dependent value of the scalar field,
which for the case at hand is $\phi\sim 3.5\,M_p$. Hence, it is
possible for this fermion-loop mechanism to end the power-law
inflationary era, without affecting significantly the power-law
behavior of the potential. One can also choose much smaller Yukawa
couplings to achieve a better behavior of the potential too, and
also appropriately choose the renormalization scale. However this
analysis exceeds by far the aims of this work, so we refrain from
going into details.

\section{Conclusions and Discussion}

In this work we analyzed quantitatively a scenario which can solve
the TCC problems of standard slow-roll inflation. Specifically we
considered an $R^2$-corrected scalar field theory with an
exponential scalar potential. This theoretical framework can
remedy the TCC problems of slow-roll inflation, if the $R^2$
gravity drives the slow-roll era and, near its end, it is followed
by a power-law inflationary era generated by the scalar field with
exponential potential. This scenario was developed in Ref.
\cite{transplanckOdintsovOikonomou} using a qualitative approach
which focused on the phenomenology and in this work we aimed to
quantitatively study the $R^2$-corrected scalar field theory with
an exponential scalar potential using a detailed phase space
analysis. We constructed an autonomous dynamical system for the
$F(R,\phi)$ system using appropriate dimensionless variables and
the field equations and we studied two solution subspaces of the
total phase space, the de Sitter subspace and the power-law
evolution subspace. The study of the de Sitter subspace validated
our approach that the $F(R)$ gravity would indeed dominate the
evolution if it is described by a quasi-de Sitter Hubble rate.
Specifically, we  showed that the trajectories in the phase space
are attracted to the same fixed points that vacuum $F(R)$ gravity
has. On the contrary, the power-law evolution cannot be realized
by the $F(R,\phi)$ system because the phase space is highly
unstable. To this end, we indicated that in order to solve the TCC
problems of standard slow-roll inflation, one needs an exact
power-law tail following the standard $R^2$-driven slow-roll era,
which can be solely realized by a scalar field. Thus, when the
curvature reaches a critical value near the end of the slow-roll
era, the $R^2$ gravity term in the $F(R,\phi)$ system must be
switched off. We presented a well-motivated phenomenological
$F(R)$ gravity model of this sort and we demonstrated that for
curvatures below a critical curvature near the end of the
slow-roll era, the cosmological system is composed by an
Einstein-Hilbert term and a scalar field with exponential
potential, which can realize the desired power-law tail of the
$R^2$ gravity slow-roll era, and thus solve the TCC problems of
the latter.

We also briefly discussed some implications of this work, having
to do with the existence of a quintessential scalar field theory
in the post inflationary era. This scalar field is also present at
late times and thus could be responsible for the dark energy era,
but we did not analyze this scenario in detail because it is out
of the context of the present work. Another interesting feature of
the scenario discussed in this work is related to the total EoS of
the system in the post-inflationary era and specifically during
the reheating, which can in principle have even observational
implications, for example in gravitational waves, see for example
\cite{Pi:2024kpw}. Also in this work we essentially considered a
quantum corrected scalar field action with the scalar field being
in its vacuum configuration, hence one can in principle consider
more general quantum corrections
\cite{Miao:2024shs,Miao:2024nsz,Miao:2024atw} and the
phenomenology of such models could give us interesting insights on
the primordial Universe. Finally, let us note that this
realization of a slow-roll era ending up to a power-law evolution
could also be realized by a constant-roll to slow-roll transition
as was performed in Ref. \cite{Odintsov:2017yud}. We aim to
discuss whether such a transition can resolve the TCC issues of
standard slow-roll inflation in a future work. This extension may
incorporate methods and fixed points that emerged from Refs.
\cite{Capozziello:1993xn,Carloni:2004kp}, which contain extensions
of the modified gravity sector, which we did not considered in
this work.

\section*{Acknowledgements}

This work was partially supported by the program Unidad de
Excelencia Maria de Maeztu CEX2020-001058-M, Spain (S.D.O). This
research has been funded by the Committee of Science of the
Ministry of Education and Science of the Republic of Kazakhstan
(Grant No. AP26194585) (S.D. Odintsov and V.K. Oikonomou).

\end{document}